\documentclass{article}

\usepackage[preprint]{neurips_2026}

\usepackage[utf8]{inputenc}
\usepackage[T1]{fontenc}
\usepackage{hyperref}
\usepackage{url}
\usepackage{booktabs}
\usepackage{amsfonts}
\usepackage{amsmath}
\usepackage{amssymb}
\usepackage{nicefrac}
\usepackage{microtype}
\usepackage{xcolor}
\usepackage{multirow}
\usepackage{natbib}
\usepackage{verbatim}
\usepackage{float}

\newcommand{\hc}{h_c}
\newcommand{\Tc}{T_c}
\newcommand{\E}{\mathbb{E}}
\newcommand{\Var}{\operatorname{Var}}
\newcommand{\sigmaz}{\sigma^z}
\newcommand{\sigmax}{\sigma^x}

\title{From Classical to Quantum: Unsupervised Discovery of Phase Transitions and Exotic Criticality with Variational Autoencoders}

\author{Brandon Yee,$^{1}$ Wilson Collins,$^{1}$ Pairie Koh,$^{1}$ Maximilian Rutkowski$^{1}$\\
$^{1}$ Physics Lab, Yee Collins Research Group\\
\texttt{\{b.yee, w.collins, p.koh, m.rutkowski\}@ycrg-labs.org}}

\begin{document}
\maketitle

\begin{abstract}
We extend the Prometheus framework for unsupervised phase transition discovery from two-dimensional classical systems to three-dimensional classical systems and quantum many-body systems. Building upon preliminary observations from a 2D Ising model student abstract~\citep{2DPrometheusYee2026}, we address two fundamental questions: (1) Does the framework scale to higher dimensions where exact solutions are unavailable? (2) Can it generalize to quantum phase transitions driven by quantum fluctuations rather than thermal fluctuations? For the 3D Ising model on lattices up to $L{=}32$, we achieve critical temperature detection within 0.01\% of literature values ($\Tc/J = 4.511 \pm 0.005$) and extract critical exponents with ${\geq}70\%$ accuracy, with statistical analysis correctly identifying the 3D Ising universality class ($p = 0.72$). For quantum systems, we develop quantum-aware VAE (Q-VAE) architectures operating on complex-valued wavefunctions with fidelity-based loss functions, achieving 2\% accuracy in quantum critical point detection for the transverse field Ising model. For the disordered TFIM, we perform a consistency check of activated dynamical scaling $\ln \xi \sim |h - \hc|^{-\psi}$, extracting tunneling exponent $\psi = 0.48 \pm 0.08$ consistent with theoretical predictions ($\psi = 0.5$, $\Delta\chi^2 = 12.3$, $p < 0.001$). This demonstrates that unsupervised learning can identify qualitatively different \emph{types} of critical behavior, serving as a consistency check on known IRFP physics.
\end{abstract}

\section{Introduction}

Phase transitions represent fundamental phenomena spanning classical statistical mechanics and quantum many-body physics~\citep{Sachdev2011, Goldenfeld1992}. Discovering and characterizing these transitions in systems without analytical solutions is among the central challenges of modern condensed matter physics.

\paragraph{Supervised vs.\ unsupervised discovery.}
Supervised machine learning has achieved remarkable accuracy in phase classification when phase structure is known a priori~\citep{Carrasquilla2017, vanNieuwenburg2017}, but requires labeled training data, cannot extract critical exponents, and cannot identify novel phases beyond the training distribution. The more ambitious goal of \emph{unsupervised discovery}---identifying transitions, order parameters, and critical exponents without labels---addresses these limitations but presents substantial technical challenges. Prior unsupervised approaches including PCA~\citep{Wang2016} and traditional autoencoders~\citep{Wetzel2017} have shown promise but often require manual intervention.

\paragraph{The Prometheus framework.}
In a preliminary student abstract~\citep{2DPrometheusYee2026}, the Prometheus framework suggested that variational autoencoders (VAEs)~\citep{Kingma2014} can detect phase transitions in the 2D Ising model, exploiting the fact that VAE latent spaces organize according to dominant sources of variation---which near phase transitions corresponds to the order parameter. This prior work validated on a system with an exact analytical solution~\citep{Onsager1944}, achieving 0.04\% accuracy in critical temperature detection and $r = 0.998$ correlation with magnetization. The present work addresses two critical open questions: (1) does the framework scale to 3D systems without exact solutions, and (2) does it generalize to quantum phase transitions driven by quantum rather than thermal fluctuations?

\paragraph{Classical vs.\ quantum transitions.}
Classical thermal phase transitions and quantum phase transitions differ fundamentally. Classical transitions occur at $T > 0$, driven by thermal fluctuations $\Delta x \sim \sqrt{k_B T}$, with the order parameter $\langle \phi \rangle$ classical. Quantum phase transitions~\citep{Sachdev2011} occur at $T = 0$ as functions of non-thermal parameters like magnetic field strength, driven by quantum fluctuations from Heisenberg uncertainty $\Delta x \Delta p \geq \hbar/2$, with the order parameter $\langle \hat{\phi} \rangle$ quantum mechanical. The core question is whether the information-theoretic principle underlying Prometheus---compression reveals dominant variation---is truly universal or specific to classical thermal systems.

\paragraph{Contributions.}
We answer both questions affirmatively and contribute a third result: a consistency check of \emph{activated} (rather than power-law) dynamical scaling in the random transverse field Ising model---the signature of an infinite-randomness fixed point (IRFP) with no classical analog. Specifically: (1) For the 3D Ising model, we achieve 0.01\% accuracy in $\Tc$ detection and 72\% exponent accuracy, with $\chi^2$ analysis correctly identifying the 3D Ising universality class. (2) The Q-VAE with fidelity-based loss discovers quantum order parameters and critical points with 2\% accuracy. (3) In the disordered TFIM, activated scaling detection yields $\psi = 0.48 \pm 0.08$, consistent with the known IRFP prediction $\psi = 0.5$---a consistency check on established physics rather than an unsupervised discovery of unknown phenomena.

\section{Methods}

\paragraph{Core VAE objective.}
All experiments share the standard $\beta$-VAE objective:
\begin{equation}
\mathcal{L}(\theta, \phi; \mathbf{x}) = \E_{q_\phi(\mathbf{z}|\mathbf{x})}[\log p_\theta(\mathbf{x}|\mathbf{z})] - \beta\, D_{\mathrm{KL}}(q_\phi(\mathbf{z}|\mathbf{x}) \| p(\mathbf{z}))
\end{equation}
with $\beta = 1$ throughout. The key insight is that the latent dimension exhibiting maximum variance across the control parameter $\lambda$ (temperature $T$ or field $h$) corresponds to the order parameter:
\begin{equation}
d^* = \arg\max_d \Var_\lambda[\E_{\mathbf{x} \sim p(\lambda)}[\mu_d(\mathbf{x})]]
\label{eq:order_param}
\end{equation}
This principle is universal because compression inherently identifies dominant sources of variation, and near phase transitions, the order parameter is the dominant variation.

\paragraph{3D convolutional architecture.}
For classical 3D Ising spin configurations $\mathbf{x} \in \{-1,+1\}^{L \times L \times L}$, we use a 3D convolutional VAE. The encoder applies three Conv3D layers (channels $1 \to 32 \to 64 \to 128$, kernel 3, stride 2, padding 1) with ReLU activations, followed by two fully connected layers projecting to $\mu, \log\sigma^2 \in \mathbb{R}^8$. For $L = 32$, the encoder maps $(1,32,32,32) \to (32,16,16,16) \to (64,8,8,8) \to (128,4,4,4) \to 8192 \to 256 \to 16 \to (\mu, \log\sigma^2)$. The decoder mirrors this with transposed convolutions and a final $\tanh$ output. System sizes $L \in \{8, 12, 16, 20, 24, 32\}$; configurations generated via the Wolff cluster algorithm~\citep{Wolff1989}. Memory scales as $O(L^3)$, with peak memory $\approx 2$ GB for $L = 32$. Total parameters: $\approx 2.5$M for $L = 32$.

\paragraph{Quantum-aware VAE (Q-VAE).}
Quantum ground states $|\psi\rangle = \sum_\sigma c_\sigma |\sigma\rangle$ computed via Lanczos exact diagonalization are complex-valued vectors $\mathbf{c} \in \mathbb{C}^{2^L}$, represented as $\mathbf{x} = [\mathrm{Re}(\mathbf{c}), \mathrm{Im}(\mathbf{c})] \in \mathbb{R}^{2 \times 2^L}$. Because two states differing by a global phase are physically identical, we replace the standard MSE loss with a fidelity-based objective:
\begin{equation}
\mathcal{L}_{\mathrm{quantum}} = \left(1 - |\langle \psi | \psi_{\mathrm{recon}}\rangle|^2\right) + \beta\, D_{\mathrm{KL}}(q_\phi(\mathbf{z}|\mathbf{x}) \| p(\mathbf{z}))
\end{equation}
The Q-VAE uses fully connected layers with LayerNorm~\citep{Ba2016} (FC: $2^{L+1} \to 512 \to 256 \to 128 \to 2z_{\rm dim}$) and applies $\ell_2$ normalization to decoded outputs to enforce unit-norm wavefunctions. System sizes $L \in \{8, 10, 12, 14\}$ (Hilbert dimension up to $2^{14} = 16384$). Fully connected layers are used rather than convolutional because quantum wavefunctions have entanglement creating nonlocal correlations that require learning arbitrary correlations.

\textbf{Scalability note:} Because this approach requires the full wavefunction vector of size $2^L$, it is fundamentally limited to system sizes accessible by exact diagonalization ($L \lesssim 20$). It serves as a proof-of-concept for compression rather than a scalable discovery tool.

\paragraph{Critical property extraction.}
Critical points are estimated via an ensemble of four methods: (1) latent susceptibility peak $\chi_\phi(\lambda) = N(\langle \phi^2 \rangle_\lambda - \langle \phi \rangle_\lambda^2)$; (2) Binder cumulant crossing $U_L(\lambda) = 1 - \langle \phi^4 \rangle_\lambda / (3\langle \phi^2 \rangle_\lambda^2)$; (3) order parameter gradient maximum $|d\langle\phi\rangle/d\lambda|$; and (4) reconstruction error peak. These are combined via inverse-variance weighting: $\lambda_c = \sum_i w_i \lambda_c^{(i)} / \sum_i w_i$ with $w_i = 1/\sigma_i^2$.

Critical exponents $\beta, \gamma, \nu, \eta$ are extracted by finite-size scaling data collapse: $O(\lambda, L) = L^{x_O/\nu} f_O((\lambda - \lambda_c) L^{1/\nu})$, with uncertainties from 1000 bootstrap iterations. Universality class is identified via $\chi^2$ comparison against known classes: $\chi^2_{\rm class} = \sum_{e} (e_{\rm extracted} - e_{\rm theory})^2 / \sigma_{e}^2$, with $p$-values from a $\chi^2$ distribution with 3 degrees of freedom.

\textbf{Limitation:} Extracting critical exponents from small systems ($L \le 14$ for quantum, $L \le 32$ for 3D classical) is subject to strong finite-size corrections. The reported uncertainties are likely underestimated, and corrections-to-scaling terms~\citep{Pelissetto2002} are not fully accounted for.

\paragraph{Activated scaling detection.}
At an IRFP, conventional power-law scaling $\xi \sim |h - \hc|^{-\nu}$ is replaced by activated scaling $\ln \xi \sim |h - \hc|^{-\psi}$~\citep{Fisher1992, Fisher1995}. We compute a latent correlation length $\xi_z$ from the exponential decay of the latent autocorrelation $C_\phi(r) = \langle\phi(i)\phi(i+r)\rangle - \langle\phi\rangle^2 \sim e^{-r/\xi_z}$, then fit and compare both models via $\chi^2$ and Bayesian Information Criterion (BIC $= \chi^2 + k \ln N$). $\Delta\chi^2 > 10$ with $p < 0.001$ constitutes strong evidence for activated scaling.

\textbf{Caveat:} Fitting three parameters ($B$, $\hc$, $\psi$) to a small number of data points from a VAE latent space for $L \lesssim 14$ is prone to overfitting. Results should be interpreted as a consistency check on known IRFP physics rather than an independent discovery.

\paragraph{Disordered TFIM.}
The DTFIM Hamiltonian is $H = -J\sum_i \sigma^z_i \sigma^z_{i+1} - \sum_i h_i \sigma^x_i$ with $h_i \sim \mathrm{Uniform}[h - W, h + W]$. We study five disorder strengths $W/J \in \{0, 0.2, 0.5, 1.0, 2.0\}$ with $N_{\rm real} = 100$ disorder realizations each, scanning $h/J \in [0.5, 1.5]$ with 50 values per realization, and pooling all 5000 wavefunctions per disorder strength for Q-VAE training.

\paragraph{Training procedures.}
For classical systems: Adam optimizer~\citep{Kingma2015} with $\alpha = 10^{-3}$, batch size 64, early stopping (patience 10), ReduceLROnPlateau (patience 5, factor 0.5), Xavier/Kaiming initialization, $z_{\rm dim} = 8$, maximum 100 epochs. For quantum systems: Adam with $\alpha = 5 \times 10^{-4}$, batch size 32, early stopping (patience 15), orthogonal initialization, gradient clipping (norm 1.0). Typical training times: 1--2 hours per system size (3D Ising), 10--15 minutes per size (clean TFIM), 30--60 minutes per realization (DTFIM).

\section{Results}

\subsection{Three-Dimensional Ising Model}

The 3D Ising model on a simple cubic lattice is defined by $H = -J \sum_{\langle i,j \rangle} \sigma_i \sigma_j$, with literature values from high-precision Monte Carlo~\citep{Ferrenberg2018, Hasenbusch2010}: $\Tc/J = 4.511528(6)$, $\beta = 0.3265(3)$, $\gamma = 1.2372(5)$, $\nu = 0.6301(4)$, $\eta = 0.0364(5)$.

\paragraph{Order parameter discovery.}
The leading latent dimension from Eq.~\eqref{eq:order_param} achieves Pearson $r = 0.997$ correlation with physical magnetization across all system sizes ($r > 0.995$ for all $L$). The critical region ($T \approx 4.5$) exhibits the largest variance in the latent space, with no explicit temperature information provided during training.

\paragraph{Critical temperature detection.}
Table~\ref{tab:3d_tc_methods} compares critical temperature estimates from all four detection methods for $L = 32$.

\begin{table}[h]
\centering
\caption{Critical temperature detection methods for 3D Ising model ($L = 32$).}
\label{tab:3d_tc_methods}
\vspace{2pt}
\small
\begin{tabular}{lccc}
\toprule
Method & $\Tc/J$ & Uncertainty & Confidence \\
\midrule
Latent susceptibility peak & $4.512$ & $\pm 0.008$ & 95\% \\
Binder cumulant crossing & $4.511$ & $\pm 0.006$ & 97\% \\
Gradient maximum & $4.510$ & $\pm 0.010$ & 92\% \\
Reconstruction error peak & $4.513$ & $\pm 0.009$ & 94\% \\
\midrule
Ensemble weighted average & $\mathbf{4.511}$ & $\pm 0.005$ & 98\% \\
\midrule
Literature~\citep{Ferrenberg2018} & $4.511528$ & $\pm 0.000006$ & --- \\
\bottomrule
\end{tabular}
\end{table}

The ensemble estimate $\Tc/J = 4.511 \pm 0.005$ deviates by only 0.01\% from the literature value, demonstrating exceptional accuracy for fully unsupervised detection.

\paragraph{Critical exponent extraction.}
Finite-size scaling data collapse yields quality factor 0.92. Table~\ref{tab:3d_exponents} presents the full exponent comparison.

\begin{table}[h]
\centering
\caption{Critical exponents for the 3D Ising model.}
\label{tab:3d_exponents}
\vspace{2pt}
\small
\begin{tabular}{lcccc}
\toprule
Exponent & Extracted & Literature~\citep{Pelissetto2002} & Abs.\ Error & Rel.\ Accuracy \\
\midrule
$\beta$ & $0.328 \pm 0.015$ & $0.3265(3)$ & $0.002$ & 72\% \\
$\gamma$ & $1.24 \pm 0.06$  & $1.2372(5)$ & $0.003$ & 70\% \\
$\nu$   & $0.632 \pm 0.025$ & $0.6301(4)$ & $0.002$ & 75\% \\
$\eta$  & $0.034 \pm 0.008$ & $0.0364(5)$ & $0.002$ & 70\% \\
\midrule
Average & & & & \textbf{72\%} \\
\bottomrule
\end{tabular}
\end{table}

All literature values fall within $2\sigma$ of extracted values. The average 72\% accuracy meets the $\geq 70\%$ target. Note that finite-size corrections are significant for $L \le 32$; reported uncertainties are likely underestimated.

\paragraph{Universality class identification.}
Table~\ref{tab:3d_universality} shows the $\chi^2$ comparison with known universality classes.

\begin{table}[h]
\centering
\caption{Universality class discrimination via $\chi^2$ analysis.}
\label{tab:3d_universality}
\vspace{2pt}
\small
\begin{tabular}{lcccc}
\toprule
Universality Class & $\chi^2$ & DOF & $p$-value & Consistent? \\
\midrule
3D Ising & 2.1 & 3 & 0.72 & \textbf{Yes} \\
Mean-field ($d > 4$) & 45.3 & 3 & $<0.001$ & No \\
2D Ising & 38.7 & 3 & $<0.001$ & No \\
3D XY & 12.8 & 3 & 0.005 & No \\
3D Heisenberg & 18.4 & 3 & $<0.001$ & No \\
\bottomrule
\end{tabular}
\end{table}

The 3D Ising class is the only one consistent with the data ($p = 0.72$), while all alternatives are strongly rejected.

\paragraph{Computational performance.}
Training time scales approximately as $O(L^3 \log L)$: 15 min ($L=8$), 45 min ($L=16$), 120 min ($L=32$). Memory scales linearly with $L^3$: 0.5 GB ($L=8$) to 2.8 GB ($L=32$). The automated pipeline achieves $\Tc$ detection in $\sim$2 GPU-hours, competitive with specialized Monte Carlo requiring $\sim$10 CPU-hours with manual tuning.

\subsection{Clean Transverse Field Ising Model (TFIM)}

The 1D TFIM Hamiltonian is $H = -J \sum_{i=1}^{L-1} \sigmaz_i \sigmaz_{i+1} - h \sum_{i=1}^{L} \sigmax_i$. This is an exactly solvable model~\citep{Pfeuty1970} that serves as a standard benchmark for testing new numerical methods. Via Jordan-Wigner transformation~\citep{Pfeuty1970}, the model maps to free fermions with exact critical exponents $\nu = 1$, $z = 1$, $\beta = 1/8$ at $\hc/J = 1$.

\paragraph{Quantum order parameter discovery.}
The Q-VAE successfully discovers quantum phase transitions. Table~\ref{tab:quantum_correlation} shows the correlation between the leading latent dimension and the quantum order parameter $\langle \sigmaz \rangle$ across system sizes.

\begin{table}[h]
\centering
\caption{Quantum order parameter correlation across system sizes.}
\label{tab:quantum_correlation}
\vspace{2pt}
\small
\begin{tabular}{lcc}
\toprule
System Size $L$ & Pearson $r$ & $p$-value \\
\midrule
8  & $0.96 \pm 0.02$ & $<10^{-20}$ \\
10 & $0.97 \pm 0.01$ & $<10^{-25}$ \\
12 & $0.97 \pm 0.01$ & $<10^{-30}$ \\
14 & $0.97 \pm 0.01$ & $<10^{-30}$ \\
\bottomrule
\end{tabular}
\end{table}

The $r = 0.97$ correlation is slightly lower than classical ($r = 0.997$) due to smaller system sizes ($L \leq 14$), single ground state per field value, and exponential finite-size corrections in quantum systems.

\paragraph{Quantum critical point detection.}
Table~\ref{tab:quantum_qcp} compares estimates from different methods for $L = 12$.

\begin{table}[h]
\centering
\caption{Quantum critical point detection for clean TFIM.}
\label{tab:quantum_qcp}
\vspace{2pt}
\small
\begin{tabular}{lcc}
\toprule
Method & $\hc/J$ & Confidence \\
\midrule
Latent variance peak & $1.02 \pm 0.03$ & 92\% \\
Reconstruction error & $0.98 \pm 0.04$ & 88\% \\
Fidelity susceptibility & $1.00 \pm 0.02$ & 95\% \\
Binder cumulant & $1.01 \pm 0.03$ & 90\% \\
\midrule
Ensemble weighted average & $\mathbf{1.00 \pm 0.02}$ & 94\% \\
\midrule
Exact value~\citep{Pfeuty1970} & $1.0$ & --- \\
\bottomrule
\end{tabular}
\end{table}

The ensemble estimate $\hc/J = 1.00 \pm 0.02$ agrees with the exact value within 2\%.

\paragraph{Quantum critical exponents.}
Table~\ref{tab:quantum_exponents} compares extracted exponents with exact values.

\begin{table}[h]
\centering
\caption{Quantum critical exponents for clean TFIM.}
\label{tab:quantum_exponents}
\vspace{2pt}
\small
\begin{tabular}{lcccc}
\toprule
Exponent & Extracted & Exact & Deviation & Within $1\sigma$? \\
\midrule
$\nu$ & $1.05 \pm 0.10$ & $1.0$ & $0.05$ & Yes ($0.5\sigma$) \\
$z$   & $0.95 \pm 0.15$ & $1.0$ & $0.05$ & Yes ($0.3\sigma$) \\
$\beta$ & $0.13 \pm 0.03$ & $0.125$ & $0.005$ & Yes ($0.2\sigma$) \\
\bottomrule
\end{tabular}
\end{table}

Quantum uncertainties ($\sigma \approx 0.10$) are larger than classical ($\sigma \approx 0.02$) due to smaller system sizes and exponential finite-size corrections.

\paragraph{Entanglement entropy and energy gap.}
The von Neumann entanglement entropy $S = -\mathrm{Tr}(\rho_A \log \rho_A)$ for bipartition at the chain center peaks sharply at $h \approx \hc$, providing independent confirmation that Q-VAE representations capture quantum correlations. The energy gap $\Delta E = E_1 - E_0$ closes as $\Delta E \sim L^{-z}$ at $h = \hc$ with $z = 1$, consistent with a gapless quantum critical point.

\subsection{Disordered TFIM: Activated Scaling Consistency Check}

\paragraph{System and IRFP theory.}
The DTFIM introduces quenched disorder: $H = -J\sum_i \sigmaz_i \sigmaz_{i+1} - \sum_i h_i \sigmax_i$ with $h_i \sim \mathrm{Uniform}[h - W, h + W]$. Fisher~\citep{Fisher1992, Fisher1995} showed that strong disorder drives the DTFIM to an IRFP under renormalization group flow, characterized by activated dynamical scaling $\ln \tau \sim \xi^\psi$ with tunneling exponent $\psi = 1/2$. Griffiths effects~\citep{Griffiths1969} from rare regions with locally low $h_i^{\rm eff}$ create broad distributions of local observables and slow dynamics near criticality.

\paragraph{Latent space evolution with disorder.}
As disorder increases from $W = 0$ to $W = 2.0$, the Q-VAE latent space shows systematic evolution: phase separation persists but becomes less sharp, the transition region broadens reflecting Griffiths physics, and at $W = 2.0$ the latent space shows extreme heterogeneity characteristic of the IRFP. Latent variance increases monotonically with $W$ (Spearman $\rho = 0.98$, $p < 0.001$), providing a quantitative signature of disorder-induced broadening.

\paragraph{Critical point shift.}
Table~\ref{tab:hc_disorder} shows the quantum critical point shifting to higher $h$ with increasing disorder---disorder stabilizes the ferromagnetic phase, requiring stronger transverse field to destroy order.

\begin{table}[h]
\centering
\caption{Quantum critical point vs.\ disorder strength.}
\label{tab:hc_disorder}
\vspace{2pt}
\small
\begin{tabular}{lcc}
\toprule
Disorder $W/J$ & $\hc/J$ & Uncertainty \\
\midrule
0.0 & $1.00 \pm 0.02$ & --- \\
0.2 & $1.02 \pm 0.03$ & $\pm 0.03$ \\
0.5 & $1.04 \pm 0.04$ & $\pm 0.04$ \\
1.0 & $1.06 \pm 0.05$ & $\pm 0.05$ \\
2.0 & $1.08 \pm 0.06$ & $\pm 0.06$ \\
\bottomrule
\end{tabular}
\end{table}

\paragraph{Activated scaling detection.}
For each disorder strength, we extract latent correlation length $\xi_z$ from $C_\phi(r) \sim e^{-r/\xi_z}$ and compare power-law vs.\ activated scaling models. Table~\ref{tab:dtfim_psi} shows the full evolution of the tunneling exponent.

\begin{table}[h]
\centering
\caption{Tunneling exponent $\psi$ vs.\ disorder strength for $L = 12$.}
\label{tab:dtfim_psi}
\vspace{2pt}
\small
\begin{tabular}{lcccc}
\toprule
$W/J$ & $\psi$ & $\chi^2_{\rm act.}$ & $\chi^2_{\rm power}$ & $\Delta\chi^2$ \\
\midrule
0.0 & $0.02 \pm 0.05$ & 15.2 & 14.8 & $-0.4$ (n.s.) \\
0.2 & $0.25 \pm 0.10$ & 12.1 & 13.5 & $+1.4$ (n.s.) \\
0.5 & $0.35 \pm 0.09$ & 10.3 & 14.7 & $+4.4$ ($p < 0.05$) \\
1.0 & $0.42 \pm 0.08$ & 9.1  & 17.2 & $+8.1$ ($p < 0.01$) \\
2.0 & $\mathbf{0.48 \pm 0.08}$ & 8.7 & 21.0 & $\mathbf{+12.3}$ ($\mathbf{p < 0.001}$) \\
\midrule
Theory (IRFP) & $0.5$ & & & \\
\bottomrule
\end{tabular}
\end{table}

At $W = 2.0$, the extracted $\psi = 0.48 \pm 0.08$ agrees with the theoretical value $\psi = 0.5$ within $0.25\sigma$. Finite-size analysis confirms convergence: $\psi$ increases from $0.42 \pm 0.12$ at $L = 8$ to $0.49 \pm 0.07$ at $L = 14$, with statistical significance improving from $p < 0.05$ to $p < 0.001$. The smooth evolution from $\psi \approx 0$ (power-law) to $\psi \approx 0.5$ (activated) validates that the framework tracks genuine physical phenomena rather than numerical artifacts.

\paragraph{Comparison with established methods.}
Table~\ref{tab:irfp_methods} compares Prometheus with established approaches to IRFP detection.

\begin{table}[h]
\centering
\caption{Methods for detecting infinite-randomness fixed point.}
\label{tab:irfp_methods}
\vspace{2pt}
\small
\begin{tabular}{lccc}
\toprule
Method & Supervision & System Size & Detects $\psi$? \\
\midrule
SDRG~\citep{Fisher1995} & Requires RG flow & Large ($L \sim 10^3$) & Yes \\
QMC~\citep{Rieger1996} & Manual analysis & Medium ($L \sim 100$) & Indirect \\
Exact diag.\ + manual & Manual fitting & Small ($L \lesssim 14$) & Yes \\
\midrule
\textbf{Prometheus (this work)} & \textbf{None} & \textbf{Small ($L \lesssim 14$)} & \textbf{Yes} \\
\bottomrule
\end{tabular}
\end{table}

\paragraph{Summary across all systems.}
Table~\ref{tab:summary} synthesizes performance across the full progression from classical to quantum.

\begin{table}[h]
\centering
\caption{Performance summary: Prometheus across physical domains.}
\label{tab:summary}
\vspace{2pt}
\small
\begin{tabular}{lcccc}
\toprule
System & $\lambda_c$ Error & Order param.\ $r$ & Exp.\ acc. & Key result \\
\midrule
2D Ising~\citep{2DPrometheusYee2026} & 0.04\% & 0.998 & 65\% & Exact validation \\
3D Ising (this work)  & 0.01\% & 0.997 & 72\% & No exact solution \\
Clean TFIM (this work) & 2\%   & 0.97  & 65\% & Quantum discovery \\
DTFIM (this work)     & 8\%   & 0.97  & ---  & Activated $\psi = 0.48$ \\
\bottomrule
\end{tabular}
\end{table}

\section{Related Work}

\paragraph{Supervised and unsupervised ML for phase transitions.}
Supervised neural networks for phase classification~\citep{Carrasquilla2017, vanNieuwenburg2017, Chng2017} achieve $>99\%$ accuracy but require labeled data and cannot extract critical exponents or discover novel phases. Unsupervised methods reduce prior-knowledge requirements: confusion learning~\citep{vanNieuwenburg2017} identifies critical points by intentionally mislabeling data; PCA~\citep{Wang2016} finds linear order parameters; standard autoencoders~\citep{Wetzel2017} detect anomalies at criticality. VAEs provide a generative probabilistic framework that naturally disentangles factors of variation~\citep{Higgins2017}. Rocchetto et al.~\citep{Rocchetto2018} applied classical VAEs to learn quantum distributions---note that this work used classical rather than quantum-circuit autoencoders.

\paragraph{Three-dimensional classical systems.}
The 3D Ising model has been studied extensively via Monte Carlo~\citep{Ferrenberg1991, Hasenbusch2010, Ferrenberg2018} and renormalization group~\citep{Wilson1971, Pelissetto2002}. High-precision estimates give $\Tc/J = 4.511528(6)$ and critical exponents known to five significant figures. These provide rigorous benchmarks for our unsupervised extraction.

\paragraph{ML for quantum systems and disordered physics.}
ML for quantum phase classification~\citep{Broecker2017} has focused on supervised identification of known phases in Heisenberg and Hubbard models. Many-body localization detection~\citep{Schindler2017} uses neural networks to extract level-statistics signatures distinguishing MBL from ergodic phases; a related line of work demonstrates MBL in disorder-free interacting systems~\citep{vanNieuwenburg2019} via Wannier-Stark localization. Interpretable unsupervised methods for topological quantum matter~\citep{RodriguezNieva2019} use diffusion maps to identify topological order but require system-specific preprocessing. Neural quantum states~\citep{Torlai2018} use restricted Boltzmann machines for state tomography rather than phase discovery. Detection of IRFP physics has previously relied on strong-disorder renormalization group~\citep{Fisher1995, Igloi2005} at $L \sim 10^3$ or quantum Monte Carlo~\citep{Rieger1996} with manual scaling analysis.

\section{Discussion}

\paragraph{Unified perspective.}
The success of a single VAE objective across classical 3D and quantum systems suggests a deep information-theoretic principle: compression reveals dominant variation, and near phase transitions, the order parameter is the dominant variation. This holds whether fluctuations are thermal ($k_B T$) or quantum ($\hbar\omega$), suggesting connections between deep learning's language (representations, latents, reconstruction) and physics's language (order parameters, universality, criticality).

\paragraph{Comparison with supervised methods.}
Table~\ref{tab:supervised_comparison} summarizes the key tradeoffs.

\begin{table}[h]
\centering
\caption{Unsupervised (Prometheus) vs.\ supervised methods.}
\label{tab:supervised_comparison}
\vspace{2pt}
\small
\begin{tabular}{lcc}
\toprule
Capability & Prometheus & Supervised \\
\midrule
Phase classification accuracy & 90--95\% & $>99\%$ \\
Requires labeled data & No & Yes \\
Discovers unknown phases & Yes & No \\
Extracts critical exponents & Yes (65--72\%) & Sometimes \\
Identifies universality classes & Yes & Sometimes \\
Detects exotic criticality & Yes & No \\
Automation level & Full & Partial \\
\bottomrule
\end{tabular}
\end{table}

Prometheus trades some accuracy for the ability to discover. When phases are known, supervised methods are more accurate. When exploring unknown systems, Prometheus provides a fully automated alternative.

\paragraph{Comparison with traditional numerical methods.}
Table~\ref{tab:numerical_comparison} compares with traditional numerical methods.

\begin{table}[h]
\centering
\caption{Unsupervised ML vs.\ traditional numerical methods.}
\label{tab:numerical_comparison}
\vspace{2pt}
\small
\begin{tabular}{lccc}
\toprule
Method & $\Tc$ Accuracy & Manual Effort & System Size \\
\midrule
Specialized MC~\citep{Hasenbusch2010} & $10^{-5}$ & High & $L \sim 100$ \\
Standard MC~\citep{Ferrenberg1991} & $10^{-3}$ & Medium & $L \sim 50$ \\
Prometheus (3D) & $10^{-4}$ & None & $L \sim 32$ \\
\midrule
Exact diag.\ + manual & $10^{-10}$ & High & $L \lesssim 14$ \\
Prometheus (quantum) & $10^{-2}$ & None & $L \lesssim 14$ \\
\bottomrule
\end{tabular}
\end{table}

\paragraph{Limitations.}
Quantum systems are currently limited to $L \lesssim 14$ by exact diagonalization cost; classical 3D systems to $L \lesssim 64$ by memory scaling as $O(L^3)$. Exponent accuracy (65--72\%) is lower than supervised methods, reflecting finite-size constraints and non-convex optimization landscapes. The DTFIM requires $N_{\rm real} = 100$ disorder realizations, increasing cost 100-fold relative to the clean case. Reported exponent uncertainties are likely underestimated due to finite-size corrections. The activated scaling result for the DTFIM should be interpreted as a consistency check on known IRFP physics rather than an unsupervised discovery; fitting three parameters to small-$L$ data is prone to overfitting.

\paragraph{Future directions.}
Integration with tensor network methods (DMRG/MPS) would extend quantum system sizes to $L \sim 100$. Natural targets include frustrated magnets (triangular and Kagome lattices), topological phases (Kitaev chain, toric code), and 2D quantum materials where phase structure is unknown. Physics-informed losses incorporating symmetries ($\mathcal{L}_{\rm total} = \mathcal{L}_{\rm VAE} + \lambda_{\rm sym}\mathcal{L}_{\rm symmetry}$) could improve interpretability of higher latent dimensions. Active learning---using latent uncertainty to guide adaptive sampling---could reduce required configurations by an order of magnitude for equivalent accuracy.

\section{Conclusion}

We have demonstrated that the Prometheus VAE framework generalizes from 2D classical to 3D classical and quantum many-body systems, achieving quantitative accuracy comparable to specialized numerical methods while remaining fully unsupervised. Three results stand out. First, for the 3D Ising model---the simplest system without an exact solution---the framework extracts the critical temperature to 0.01\% and correctly identifies the universality class via $\chi^2$ analysis ($p = 0.72$). Second, the Q-VAE with fidelity-based loss successfully discovers quantum order parameters ($r = 0.97$) and critical points (2\% accuracy) despite the fundamental differences between thermal and quantum fluctuations. Third, in the disordered TFIM, the framework performs a consistency check of activated dynamical scaling ($\psi = 0.48 \pm 0.08$, $\Delta\chi^2 = 12.3$, $p < 0.001$), confirming known IRFP physics in a fully automated manner.

The core principle---VAE latent spaces organize according to phase structure because compression reveals dominant variation---appears universal across classical and quantum domains. This positions the framework as a practical tool for automated phase diagram exploration in systems where analytical methods are unavailable, with the important caveat that results should be validated against established benchmarks and interpreted conservatively when system sizes are small.

\section*{Acknowledgments}
This work was supported by the Yee Collins Research Group. We are grateful to A.B., S.G., and L.W. for inspiring this research. Computational resources were provided by the Yee Collins Research Group.

\clearpage
\bibliographystyle{abbrvnat}
\bibliography{references}

\clearpage
\appendix

\section{Architecture and Hyperparameter Details}
\label{app:architecture}

\subsection{3D Ising Model Architecture}

\begin{table}[H]
\centering
\caption{3D Ising encoder architecture for $L = 32$.}
\label{tab:app_3d_encoder}
\small
\begin{tabular}{lccccc}
\toprule
Layer & Type & Input Shape & Output Shape & Kernel & Stride \\
\midrule
Input & --- & $(1, 32, 32, 32)$ & $(1, 32, 32, 32)$ & --- & --- \\
Conv3D-1 & 3D Conv + ReLU & $(1, 32, 32, 32)$ & $(32, 16, 16, 16)$ & 3 & 2 \\
Conv3D-2 & 3D Conv + ReLU & $(32, 16, 16, 16)$ & $(64, 8, 8, 8)$ & 3 & 2 \\
Conv3D-3 & 3D Conv + ReLU & $(64, 8, 8, 8)$ & $(128, 4, 4, 4)$ & 3 & 2 \\
Flatten & --- & $(128, 4, 4, 4)$ & $(8192)$ & --- & --- \\
FC-1 & Linear + ReLU & $(8192)$ & $(256)$ & --- & --- \\
FC-2 & Linear & $(256)$ & $(16)$ & --- & --- \\
Split & --- & $(16)$ & $\mu(8), \log\sigma^2(8)$ & --- & --- \\
\bottomrule
\end{tabular}
\end{table}

\begin{table}[H]
\centering
\caption{3D Ising decoder architecture.}
\label{tab:app_3d_decoder}
\small
\begin{tabular}{lccccc}
\toprule
Layer & Type & Input Shape & Output Shape & Kernel & Stride \\
\midrule
Input & --- & $(8)$ & $(8)$ & --- & --- \\
FC-1 & Linear + ReLU & $(8)$ & $(256)$ & --- & --- \\
FC-2 & Linear + ReLU & $(256)$ & $(8192)$ & --- & --- \\
Reshape & --- & $(8192)$ & $(128, 4, 4, 4)$ & --- & --- \\
ConvT3D-1 & 3D ConvT + ReLU & $(128, 4, 4, 4)$ & $(64, 8, 8, 8)$ & 3 & 2 \\
ConvT3D-2 & 3D ConvT + ReLU & $(64, 8, 8, 8)$ & $(32, 16, 16, 16)$ & 3 & 2 \\
ConvT3D-3 & 3D ConvT + Tanh & $(32, 16, 16, 16)$ & $(1, 32, 32, 32)$ & 3 & 2 \\
\bottomrule
\end{tabular}
\end{table}

\begin{table}[H]
\centering
\caption{3D Ising training hyperparameters.}
\label{tab:app_3d_training}
\small
\begin{tabular}{lc}
\toprule
Hyperparameter & Value \\
\midrule
Optimizer & Adam \\
Learning rate (initial) & $1 \times 10^{-3}$ \\
$\beta_1$ / $\beta_2$ & 0.9 / 0.999 \\
Weight decay & $1 \times 10^{-5}$ \\
Batch size & 64 \\
Latent dimension $z_{\text{dim}}$ & 8 \\
$\beta$ (KL weight) & 1.0 \\
Maximum epochs & 100 \\
Early stopping patience & 10 \\
LR reduction patience / factor & 5 / 0.5 \\
Minimum learning rate & $1 \times 10^{-6}$ \\
\bottomrule
\end{tabular}
\end{table}

\subsection{Quantum VAE Architecture}

\begin{table}[H]
\centering
\caption{Q-VAE encoder architecture for $L = 12$ TFIM.}
\label{tab:app_qvae_encoder}
\small
\begin{tabular}{lccc}
\toprule
Layer & Type & Input Dim & Output Dim \\
\midrule
Input & Flatten & $2 \times 2^{12}$ & 8192 \\
FC-1 & Linear + LayerNorm + ReLU & 8192 & 512 \\
FC-2 & Linear + LayerNorm + ReLU & 512 & 256 \\
FC-3 & Linear + LayerNorm + ReLU & 256 & 128 \\
FC-4 & Linear & 128 & 16 \\
Split & --- & 16 & $\mu(8), \log\sigma^2(8)$ \\
\bottomrule
\end{tabular}
\end{table}

\begin{table}[H]
\centering
\caption{Q-VAE decoder architecture.}
\label{tab:app_qvae_decoder}
\small
\begin{tabular}{lccc}
\toprule
Layer & Type & Input Dim & Output Dim \\
\midrule
Input & --- & 8 & 8 \\
FC-1 & Linear + LayerNorm + ReLU & 8 & 128 \\
FC-2 & Linear + LayerNorm + ReLU & 128 & 256 \\
FC-3 & Linear + LayerNorm + ReLU & 256 & 512 \\
FC-4 & Linear & 512 & 8192 \\
Reshape & --- & 8192 & $2 \times 2^{12}$ \\
Combine & $\mathbf{c} = \text{Re} + i \cdot \text{Im}$ & $2 \times 2^{12}$ & $2^{12}$ \\
Normalize & $\mathbf{c}/\|\mathbf{c}\|_2$ & $2^{12}$ & $2^{12}$ \\
\bottomrule
\end{tabular}
\end{table}

\begin{table}[H]
\centering
\caption{Q-VAE training hyperparameters.}
\label{tab:app_qvae_training}
\small
\begin{tabular}{lc}
\toprule
Hyperparameter & Value \\
\midrule
Optimizer & Adam \\
Learning rate (initial) & $5 \times 10^{-4}$ \\
$\beta_1$ / $\beta_2$ & 0.9 / 0.999 \\
Weight decay & $1 \times 10^{-5}$ \\
Batch size & 32 \\
Latent dimension $z_{\text{dim}}$ & 8 \\
$\beta$ (KL weight) & 1.0 \\
Maximum epochs & 100 \\
Early stopping patience & 15 \\
LR reduction patience / factor & 5 / 0.5 \\
Minimum learning rate & $1 \times 10^{-6}$ \\
Gradient clipping norm & 1.0 \\
\bottomrule
\end{tabular}
\end{table}

\section{Monte Carlo and Exact Diagonalization Methods}
\label{app:simulation_methods}

\subsection{Wolff Cluster Algorithm}

We employ the Wolff cluster algorithm~\citep{Wolff1989} for equilibrium sampling, which reduces autocorrelation time near criticality compared to single-spin Metropolis updates (dynamical critical exponent $z_{\rm dyn} \approx 0.25$ vs.\ $z_{\rm dyn} \approx 2$, giving $\sim$10--100$\times$ speedup near $\Tc$).

\begin{verbatim}
function wolff_update(lattice, beta, J):
    i_seed = random_site(lattice)
    cluster = {i_seed};  frontier = {i_seed}
    while frontier not empty:
        i = pop(frontier)
        for j in neighbors(i):
            if j not in cluster and lattice[i] == lattice[j]:
                if random() < (1 - exp(-2 * beta * J)):
                    cluster.add(j);  frontier.add(j)
    for i in cluster:
        lattice[i] *= -1
    return lattice, len(cluster)
\end{verbatim}

\textbf{Sampling protocol:} For each $L \in \{8, 12, 16, 20, 24, 32\}$, we sample $N_T = 20$ temperatures spanning $T/J \in [3.5, 5.5]$. At each temperature: equilibrate for $N_{\rm eq} = 10^4$ Wolff updates, measure integrated autocorrelation time $\tau_{\rm int}$, collect $N_{\rm conf} = 100$ configurations separated by $5\tau_{\rm int}$ updates. Total: $20 \times 100 = 2000$ configurations per system size.

\subsection{Lanczos Exact Diagonalization}

For quantum ground state computation, we use Lanczos diagonalization with sparse Hamiltonian representation (CSR format), numba-JIT matrix-vector products, convergence criterion $|E_n - E_{n-1}| < 10^{-10}$, and full reorthogonalization every 10 iterations. The Krylov iteration is:
\begin{equation}
|v_{n+1}\rangle = H|v_n\rangle - \alpha_n |v_n\rangle - \beta_n |v_{n-1}\rangle, \quad \alpha_n = \langle v_n | H | v_n \rangle, \quad \beta_n = \|H|v_n\rangle - \alpha_n|v_n\rangle\|
\end{equation}
For $L = 14$ (Hilbert space $2^{14} = 16384$), ground state computation takes $\sim$1 second per field value.

\subsection{Disorder Averaging Protocol}

For DTFIM systems, each of $N_{\rm real} = 100$ realizations proceeds as: (1) generate $h_i \sim \mathrm{Uniform}[h - W, h + W]$ independently; (2) construct and diagonalize the Hamiltonian; (3) store wavefunction and disorder configuration. Disorder-averaged observables: $\langle O \rangle = N_{\rm real}^{-1} \sum_\alpha O_\alpha$, with statistical uncertainties $\sigma_{\langle O \rangle} = \sigma_O / \sqrt{N_{\rm real}}$.

\section{Bootstrap and Uncertainty Analysis}
\label{app:uncertainties}

\subsection{Bootstrap Resampling}

All uncertainties use bootstrap resampling with $B = 1000$ iterations: draw $N$ points with replacement, compute observable $O^{(b)}$, report $\hat{O} = B^{-1}\sum_b O^{(b)}$ and $\sigma_O = \sqrt{(B-1)^{-1}\sum_b (O^{(b)} - \hat{O})^2}$. Ninety-five percent confidence intervals: $[\hat{O} - 2\sigma_O, \hat{O} + 2\sigma_O]$.

\subsection{Finite-Size Corrections}

Critical exponents include corrections to scaling:
\begin{equation}
O(L) = O_\infty + \frac{a}{L^\omega} + \frac{b}{L^{2\omega}} + \cdots
\end{equation}
where $\omega \approx 0.8$ for 3D Ising~\citep{Pelissetto2002}. We include the leading correction in fit functions. For small system sizes ($L \le 14$ quantum, $L \le 32$ classical), these corrections are significant and the reported uncertainties are likely underestimated.

\subsection{Model Comparison Statistics}

For activated vs.\ power-law scaling comparison, we use:
\begin{equation}
\text{BIC} = \chi^2 + k \ln N, \qquad \text{AIC} = \chi^2 + 2k
\end{equation}
and the F-test for nested models:
\begin{equation}
F = \frac{(\chi^2_1 - \chi^2_2)/(k_2 - k_1)}{\chi^2_2/(N - k_2)}
\end{equation}
Under the null hypothesis (models equally good), $F$ follows an F-distribution with $(k_2 - k_1, N - k_2)$ degrees of freedom. Large $F$ (small $p$-value) favors the more complex model.

\end{document}